\documentclass[aps,pre,nofootinbib,showpacs,twocolumn]{revtex4-1}%
\usepackage{amssymb,latexsym,mathrsfs}

\usepackage{amsfonts}
\usepackage{mathrsfs}
\usepackage{amsmath}
\usepackage{graphicx}
\usepackage{color}
\usepackage{hyperref}

\newcommand{\beq}{\begin{equation}}
\newcommand{\eeq}{\end{equation}}
\newcommand{\beqn}{\begin{eqnarray}}
\newcommand{\eeqn}{\end{eqnarray}}
\newcommand{\bearr}{\begin{array}}
\newcommand{\enarr}{\end{array}}

\def\bea{\begin{eqnarray}}
\def\eea{\end{eqnarray}}
\def\ba{\begin{array}}
\def\ea{\end{array}}

\begin{document}
\title{Thermal transport in the Fermi-Pasta-Ulam model with long-range interactions}
\author{Debarshee Bagchi}
\email[E-mail address:]{debarshee@cbpf.br}
\affiliation{Centro Brasileiro de Pesquisas Fisicas, Rio de Janeiro-RJ, Brazil}
\date{\today}

\begin{abstract}
We study the thermal transport properties of the one dimensional Fermi-Pasta-Ulam model ($\beta$-type) with long-range interactions.
The strength of the long-range interaction decreases with the (shortest) distance between the lattice sites as ${distance}^{-\delta}$,
where $\delta \ge 0$.
Two Langevin heat baths at unequal temperatures are connected to the ends of the one dimensional lattice via short-range harmonic
interactions that drive the system away from thermal equilibrium. In the nonequilibrium steady state the heat current, thermal
conductivity and temperature profiles are computed by solving the equations of motion numerically. 
It is found that the conductivity $\kappa$ has an interesting non-monotonic dependence with $\delta$ with a maximum at $\delta =  2.0$
for this model. Moreover, at $\delta = 2.0$, $\kappa$ diverges almost linearly with system size $N$ and the temperature profile has a
negligible slope, as one expects in ballistic transport for an integrable system.
We demonstrate that the non-monotonic behavior of the conductivity and the nearly ballistic thermal transport at $\delta = 2.0$ obtained
under nonequilibrium conditions can be explained consistently by studying the variation of largest Lyapunov exponent $\lambda_{max}$ with
$\delta$, and excess energy diffusion in the equilibrium microcanonical system.
\end{abstract}

\pacs{}

\maketitle


\section{Introduction}
\label{introduction}
Long-range interactions are ubiquitous in nature, the most common examples being gravitation and Coulomb interaction. Over the last
several decades long-ranged systems have been extensively studied and it is now known that such systems have a very rich, and often
intriguing, thermo-statistical behavior (for reviews on long-range interacting systems see \cite{LR_Levin,LR_Mukamel,LR_Ruffo}).
These systems exhibit breakdown of ergodicity, suppression of chaos, inequivalence of ensembles, long lived non-Gaussian quasi-stationary
states, phase transitions in one dimension, negative specific heat, to name a few, and therefore often possess
properties that deviate fantastically from ``well behaved'' short-ranged equilibrium systems. Needless to say, even after a lot of
effort an in-depth understanding of long-range interacting systems seems to be lacking.

In non-equilibrium statistical physics, a profusely investigated topic is thermal transport, particularly in low dimensional microscopic
models \cite{HT_Dhar,HT_LLP}. A flurry of research started after it was discovered that many simple one-dimensional (1D) models
violate the celebrated Fourier's law of heat conduction $J = -\kappa \nabla T$, where $\kappa \sim N^0$ is the constant of thermal
conductivity ($N$ being the system size), and the heat current $J$ is proportional to the temperature gradient $\nabla T$ which is
assumed to be small.
Subsequently a large number of works demonstrated that $\kappa$ is actually not a constant independent of $N$ but often diverges as
$\kappa \sim N^{\alpha}$, $0 < \alpha < 1$ \cite{HT_Dhar}. These models are frequently momentum-conserving systems i.e., without
external pinning potentials (a notable exception is the coupled rotor model \cite{Rotor}). One such model is the celebrated
Fermi-Pasta-Ulam (FPU) model that has been studied extensively in the last 50 years in statistical mechanics, chaos theory,
nonlinear dynamics and several other contexts \cite{FPUoriginal,FPU50yr,FPUparadox,FPUproblem}.

In this paper we wish to study thermal transport properties of the FPU model in 1D which has been appropriately modified to include long-range
interactions. Transport studies of long-ranged systems are few (see \cite{LRDiode,LRXYTrans,LRRect,LRTrans} for some recent works)
but are extremely important. For example, studies of thermal rectification \cite{ThDiode} in short-ranged systems show that rectification
generally ceases to exist as one approaches the thermodynamic limit \cite{Size}. This raises a question on their applicability in the
fabrication of real thermal diodes which are thought to be one of the crucial components of phononics \cite{Phononics}. In \cite{LRRect, LRDiode},
the authors studied thermal rectification properties of a long-ranged nonlinear lattice model and made an important discovery that rectification
is enhanced and also possibly survives even in the thermodynamic limit in the presence of long-range interactions. Thus, besides the purely
theoretical interest of understanding the physics of long-ranged interacting systems and low-dimensional thermal transport, these studies also
are technologically important. This is more so since presently it is possible to fabricate actual materials with long-range interactions,
such as the Coulomb crystal \cite{crystal}, the Ising pyrochlore magnets ${\rm Dy_2Ti_2O_7}$ and ${\rm Ho_2Ti_2O_7}$ \cite{Mat1,Mat2} and Permalloy
nanomagnets \cite{Mat3} with long-range magnetic interactions.

The remainder of the paper is organized as follows. In Sec. \ref{model} we describe the Fermi-Pasta-Ulam model with long-range interactions
and the numerical scheme that has been employed to study its transport properties. We describe the numerical computation of our quantities
of interest such as the heat current, conductivity, and temperature profiles. In Sec. \ref{results}, we present the result of our non-equilibrium
molecular dynamics simulation. Next, in Sec. \ref{equi}, the equilibrium (microcanonical) version of the model is studied and by computing
the largest Lyapunov exponent and spatiotemporal excess energy correlation function we explain the thermal transport features of the model
under non-equilibrium conditions. We finally summarize our main results in Sec. \ref{conclusion} and conclude with a discussion.

\section{The long-ranged FPU model with heat baths}
\label{model}


\begin{figure}[htb]
\centerline
{\includegraphics[width=8cm,angle=0]{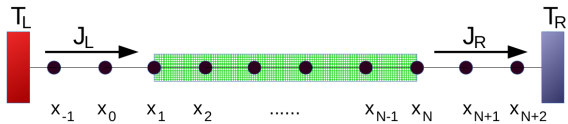}}
\caption{(Color online) Schematic diagram of the model under nonequilibrium conditions. The oscillators in the shaded (green) part
of the lattice interact via the long-range interactions obeying Eq. (\ref{H}). The oscillators at the lead sites $x_{-1}, x_0$
on the left and $x_{N+1}, x_{N+2}$ on the right interact with short-range harmonic potential $V(x) = x^2/2$ and connect the long-range
part of the lattice to the Langevin heat baths at the two ends with temperatures $T_L$ and $T_R$. The heat current that enters (exits)
the system through the left (right) end is denoted by $J_L (J_R)$ (see text).}
\label{fig:model}
\end{figure}

We consider a one-dimensional lattice of $N$ anharmonic oscillators, each with mass $m$, displacement $x_i$ and momentum $p_i$
($1 \le i \le N$) which interact with each other by long-range interaction. The Hamiltonian of our long-ranged Fermi-Pasta-Ulam (LR-FPU)
model is \cite{FPU1D,BagchiTsallis}
\begin{equation}
\mathcal{H} = \sum_i \frac{p_i^2}{2m} + \frac 12 \sum_i (x_{i+1} - x_i)^2 + \frac{1}{4 \tilde N} \sum_{i,{j \neq i}} \frac {(x_j - x_i)^4}{{|i-j|}^{\delta}}.
\label{H}
\end{equation}
Thus the interaction potential has two parts - the second term on the right in Eq. (\ref{H}) is the nearest neighbor (harmonic) interaction
whereas, the third term corresponds to the long-ranged anharmonic interactions.
The strength of the long-range interaction decays as a power-law $|i-j|^{-\delta}$ ($\delta \ge 0$) with the (shortest) distance between the
lattice sites $i$ and $j$. Thus $\delta$ sets the range of the interaction; for $\delta = 0$ we have a mean-field scenario where every
oscillator interacts with all others with equal strength, whereas for $\delta \to \infty$ the third term reduces to nearest neighbor
anharmonic interactions. Note that the prefactor $1/\tilde N$ makes the Hamiltonian Eq. (\ref{H}) extensive for all values of $\delta$
and is given as
\begin{equation}
\tilde N = \frac 1N \sum_i \sum_{j \neq i} {|i-j|}^{-\delta}.
\end{equation}
For the mean-field ($\delta = 0$) scenario $\tilde N = N$ which is the so called Kac scaling factor. Thus, inside the shaded (green) box in
Fig. \ref{fig:model}, every oscillator interacts with all others obeying Eq. (\ref{H}) with interaction strengths depending on the value of
$\delta$.

In order to study thermal transport in this kind of model we do the following alterations, as in Ref. \cite{LRDiode}. We attach $n$ additional
oscillators (we set $n = 2$, as shown in Fig. \ref{fig:model}) to both sides of the one dimensional chain and connect Langevin heat baths \cite{HT_Dhar}
at the two ends. We refer to these oscillator sites, having displacements denoted by $x_{-1}, x_0$ at the left and $x_{N+1}, x_{N+2}$ at the right 
in Fig. \ref{fig:model},
as the lead sites (or simply leads) that connect the main long-ranged system under investigation to the heat baths. The leads interact with each
other and the LR-FPU system {\it harmonically} i.e., $V(x) = x^2/2$ (spring constant set to unity). The two heat baths are set to unequal
temperatures $T_L$ and $T_R$ (we set $T_L > T_R$ always) which will drive a heat current though the long-ranged system.

The advantage of attaching the lead sites at the two ends is the ease with which one can measure the heat current $J$. Since the interactions
$\mathcal{H}$ in the main long-ranged system is quite complicated, computing the current in the bulk of the system can be cumbersome. However
exploiting the fact that when the system (including the oscillators at the lead sites) attains a nonequilibrium steady state (NESS), the current
$J_i$ at the $i$th site on the lattice is constant and independent of the lattice site index $i$ i.e., $J_i = J = constant$. Thus, one can measure
the steady state current $J$ at the lead sites to compute the heat current propagating through the main system. Since the lead site oscillators interact
with a simple potential $V(x) = x^2/2$, the current at these sites has a simple expression
\begin{equation}
J_i = - \langle \frac 12 (\dot x_{i+i} + \dot x_i) (x_{i+1} - x_i) \rangle,
\label{J}
\end{equation}
where $i$ is the site index of one of the leads.
The steady state local temperature profile can be computed using the equipartition theorem (with Boltzmann constant $k_B = 1$)
\begin{equation}
T_i = \langle m {\dot x}_i^2 \rangle
\label{Ti}
\end{equation}
and assuming that the long-ranged system achieves local thermal equilibrium at temperature $T_i$. The finite size conductivity $\kappa$
for a system of size $N$ is computed using the Fourier's law
\begin{equation}
\kappa = \frac {J N} {T_L - T_R}.
\label{k}
\end{equation}

To numerically integrate the equations of motion, we employ the second order velocity-Verlet algorithm with small time-step $\sim 10^{-2}$.
For long-ranged systems, a naive force calculation is extremely expensive computationally because of the $\mathcal{O}(N^2)$ operations
that have to be performed at each time step. This part can be accelerated by a technique which exploits the convolution theorem and using
efficient fast Fourier transformations (see \cite{SpinIce} for implementation algorithms in periodic and open spin systems with dipolar
interactions). This drastically reduces the computation time with only $\mathcal{O}(N log N)$ operations to perform at each iteration.

Starting for random initial conditions ($x_i$'s are chosen from a uniform distribution and $p_i$'s from a Gaussian distribution, both
centered at zero) we evolve the system by integrating the equations of motion until a NESS is attained.
Thereafter, we compute the heat current $J$, local temperature $T_i$ profiles, and conductivity $\kappa$ using Eqs. (\ref{J}), (\ref{Ti}),
and  (\ref{k}), respectively. We present our results obtained from the nonequilibrium simulation in the next section.

%
%

\section{Results from nonequilibrium molecular dynamics}
\label{results}
We choose the Langevin heat baths to have temperatures $T_L = T_0 + {\Delta T}/2$ and $T_R = T_0 - {\Delta T}/2$ with small
temperature difference $\Delta T$ between the two ends; thus the average temperature of the system is $\frac 12 {(T_L+T_R)} = T_0$.
Unless mentioned otherwise, we set $T_0 = 1.0$ and $\Delta T = 0.2$ ($T_L = 1.1$ and $T_R = 0.9$) for our simulations.
We employ free boundary conditions for the oscillators at the ends, although we expect our results to remain practically
unchanged even with fixed boundary conditions.

We evolve the system for a large number of iterations ($\sim 10^6$) and monitor the heat currents entering the system through the
left $J_L$ and leaving the system from the right $J_R$ (as in Fig. \ref{fig:model}). When a NESS is achieved one should obtain
$J_L = J_R$ in magnitude; $J_L,J_R$ are obtained using Eq. (\ref{J}). As an additional check for stationarity, we also computed
the velocity distribution of the individual oscillators and obtained a distribution that remains practically unaltered at large
times and, as anticipated, fits to a Gaussian distribution.
We thereafter compute the heat current as $J = \frac 12 {(J_L+J_R)}$, temperature profile $T_i$ using Eq. \ref{Ti} (with $m = 1$),
and conductivity $\kappa$ using Eq. (\ref{k}), typically for $\sim 10^8-10^9$ time iterations depending on the value of $N$ chosen.
\begin{figure}[htb]
\centerline
{\includegraphics[width=5.5cm,angle=-90]{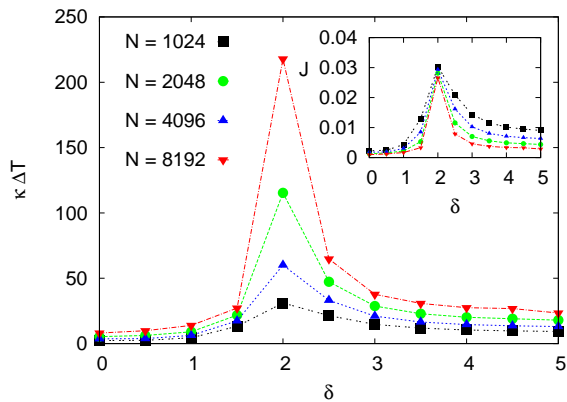}}
\caption{(Color online) Conductivity $\kappa$ (scaled by a constant $\Delta T = 0.2$) as a function of $\delta$ for different
system sizes $ N = 1024, 2048, 4096, 8192 $. The inset shows the current $J = \kappa \Delta T/N$ with $ \delta $ for average
temperature $T_0 = 1.0$. The broken lines are a guide to the eye.}
\label{fig:Curr_a}
\end{figure}
In Fig. \ref{fig:Curr_a}, we show the thermal conductivity $\kappa$ ($\Delta T = 0.2$ is a constant) of the LR-FPU model as
a function of $\delta$ for system sizes $N = 1024, 2048, 4096, 8192$ It is found that the conductivity $\kappa$ exhibits an interesting
non-monotonic behavior with $\delta$.
For small values of $\delta < 2$, $\kappa$ increases monotonically, attains a maximum value at $\delta = 2.0$ and thereafter decreases
monotonically. Thus very long-ranged  ($\delta \to 0$) as well as nearest neighbor ($\delta \to \infty$) systems have a low thermal
conductivity and maximum conductivity is obtained for an optimal $\delta = 2.0$. 
Thus, by tuning the long-range parameter $\delta$, one can manipulate the conductivity of the system over a wide range (roughly two orders of
magnitude for $N = 8192$ in Fig. \ref{fig:Curr_a}). In the inset of Fig. \ref{fig:Curr_a} the heat current $J = \kappa \Delta T/N$ is shown as
a function of $\delta$.
This is, in some sense, counter-intuitive since the range (or strength) of the long-range interaction decreases smoothly with increasing
$\delta$ and thus naively one would expect the heat current to decrease monotonically with $\delta$.
A non-monotonicity at $\delta = 1$ would also have been easy to understand, since for $\delta$ equal to the {\it embedding dimension}, these
models go from being a {\it true} long-ranged system to a short (finite) ranged system \cite{LR_Ruffo,LR_Levin}.
This in many cases, such as in the LR-FPU \cite{FPU1D} 
\footnote{The numerical result of \cite{FPU1D} in the ``weak chaos'' regime at large $N$ exhibits an unexplained saturation effect, 
(see Fig. 1a in \cite{FPU1D}, which was also later pointed out in Fig. 5(b) of \cite{BagchiTsallis}). This makes the result in \cite{FPU1D} less
reliable for large $N$ and small $\delta$. Also, here we perform our simulations for very different parameter values than what are used
in \cite{FPU1D}.},
quartic LR-FPU \cite{BagchiTsallis}, and LR-XY (coupled planar rotors) \cite{XYChaos,LLExpXY1} models, is marked by a `transition'
from weak chaos to strong chaos at $\delta = 1$ (for 1D microcanonical systems).
Instead, in the present case, it is quite intriguing that $\delta = 2.0$ appears to be a special value having maximum conductivity $\kappa$.

\begin{figure}[htb]
\centerline
{\includegraphics[width=5cm,angle=-90]{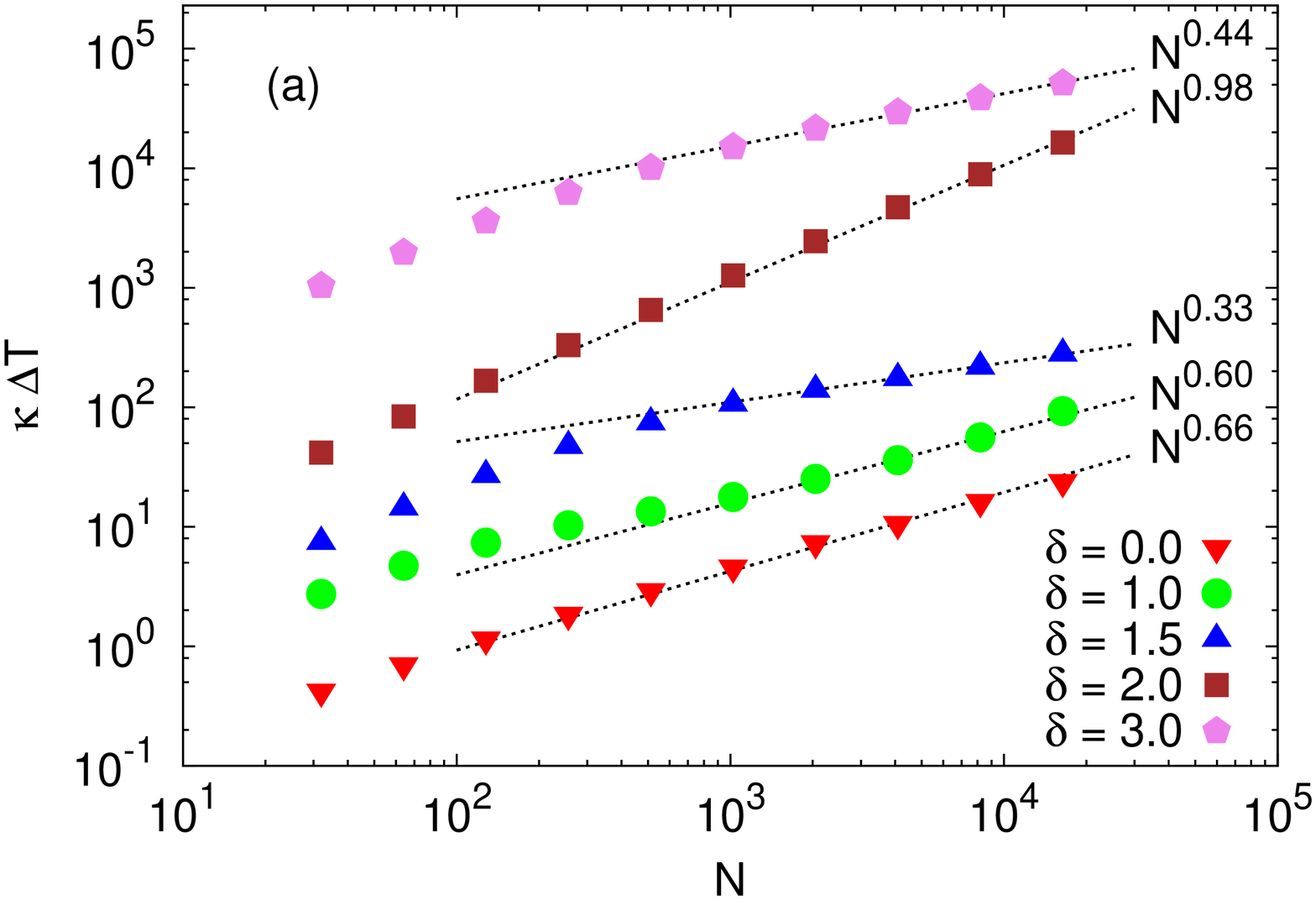}}
{\includegraphics[width=4.cm,angle=-90]{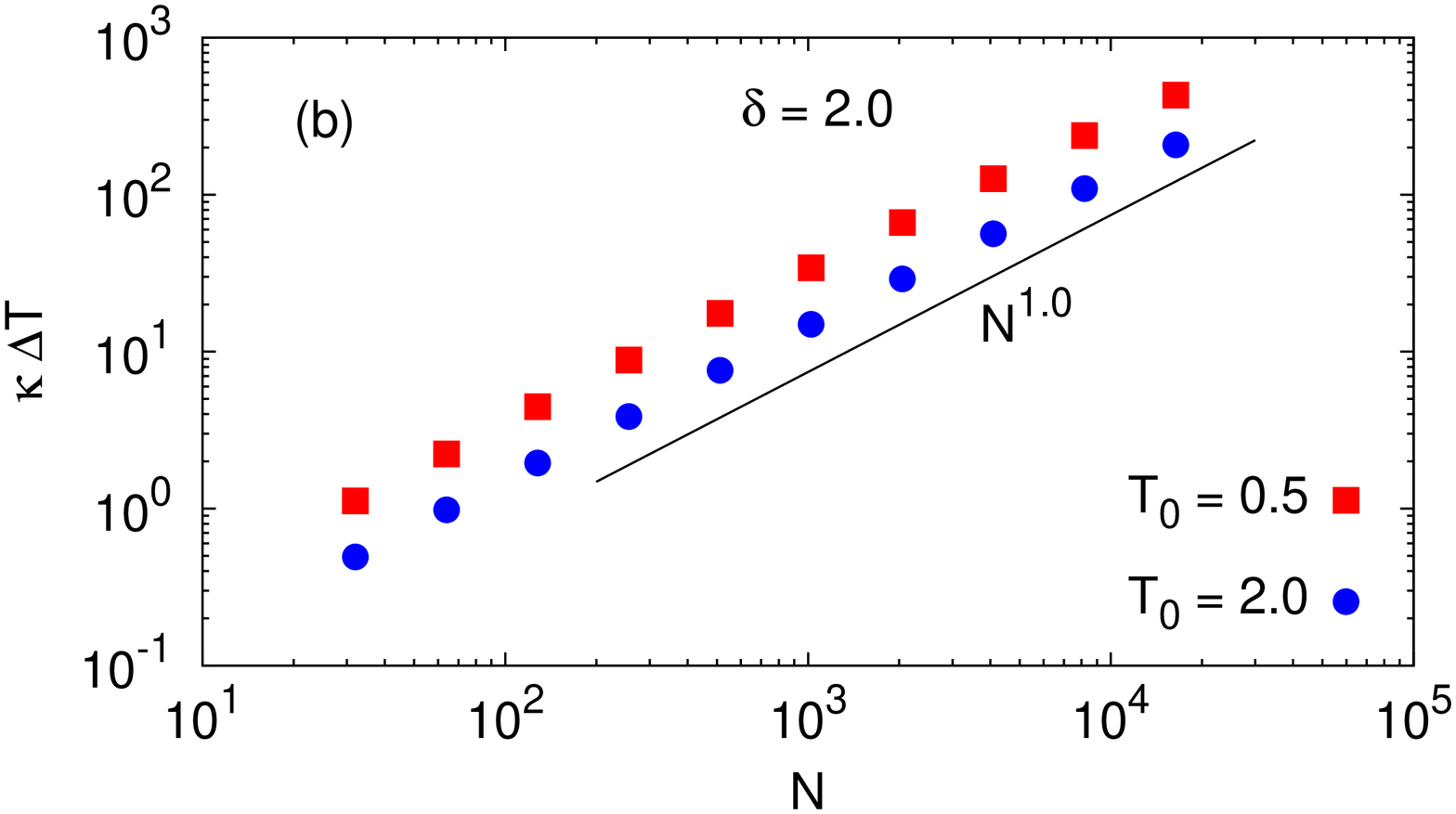}}
\caption{(Color online) (a) Log-log plot of the size dependence of conductivity $\kappa$ for different values of $\delta$. For each
$\delta$ the conductivity diverges as $\kappa \sim N^{\alpha}$. The exponent $\alpha$ for each value of $\delta$ is shown besides each
curve. (b) Log-log plot of size variation of $\kappa$ for $\delta = 2.0$ and two different average temperatures $T_0 = 0.5, 2.0$ (same
$\Delta T = 0.2$). The straight line with slope unity is shown for comparison. The curves in both (a) and (b) have been rescaled along
the $y$-direction for better visualization.}
\label{fig:kN}
\end{figure}

Next, in Fig. \ref{fig:kN}(a) we present the conductivity $\kappa$ as a function of the system size $N$ for different values of $\delta$.
We find that for all values of $\delta$ (computed up to very large values of $N = 16384$) the conductivity $\kappa$ shows a power-law
divergence $\kappa \sim N^{\alpha}$ at large $N$, as it should for momentum conserving models. However $\alpha$ changes quite drastically
(and nonmonotonically) with $\delta$, as can be clearly seen in Fig. \ref{fig:kN}(a). Surprisingly enough, corresponding to $\delta = 2.0$,
the conductivity diverges almost linearly ($\alpha \approx 0.98$) with the system size $N$ akin to ballistic heat transport in integrable
models e.g., the coupled harmonic oscillators \cite{Harmonic}.

For $\delta = 2.0$, we have checked the $N$-divergence of $\kappa$ with two other average temperatures $T_0 = 0.5, 2.0$ (and same $\Delta T = 0.2$)
and found that the exponent remains practically unaltered as shown in Fig. \ref{fig:kN}(b). We should mention that the size dependence of $\kappa$
for the LR-FPU model seems to be very different from the results for the LR-XY model studied recently \cite{LRXYTrans}. As an example, they obtain
a thermal insulator i.e., $\kappa(N) \to 0$ as $N \to \infty$ for $\delta = 0$ (Fig. 2 in \cite{LRXYTrans}), unlike what we obtain here.

\begin{figure}[htb]
\centerline
{\includegraphics[width=6cm,angle=-90]{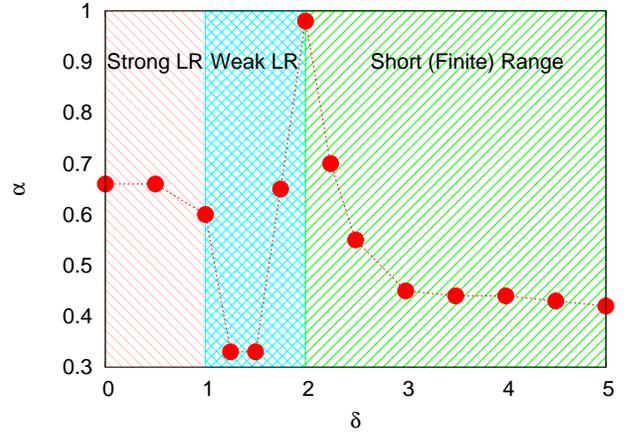}}
\caption{(Color online) Dependence of the exponent $\alpha$ on $\delta$. The parameters here are the same as in Fig. \ref{fig:kN}. The dotted line
is a guide to the eye. The three regions (from left to right) correspond to strong long-range ($0 < \delta < 1$), weak long-range ($1 < \delta < 2$)
and short(finite) range  ($\delta > 2$) interactions.}
\label{fig:alpha}
\end{figure}

In Fig. \ref{fig:alpha}, $\alpha$ is shown as a function of $\delta$ and an interesting picture emerges. We find that the
$\alpha -\delta$ plane in Fig. \ref{fig:alpha} can be split into the following regions:

(i) A region of strong long-range interaction for $0 < \delta < 1$. This region probably can be further split into two sub-regimes,
namely $0 \le \delta < 1/2$ where $\alpha \approx 0.66$ and $1/2 < \delta < 1$ for which $\alpha$ shows a decreasing behavior. A similar
suggestion has been put forward in some works \cite{LLExpXY1,dby2} since these two sub-regimes often exhibit different properties.

(ii) An intermediate region of weak long-range interaction for $1 < \delta < 2$ for which the $\alpha$ varies quite drastically
from $0.3 \lesssim \alpha \lesssim 1$.

(iii) A region of short-range (or finite range but not nearest neighbor) interactions for $\delta > 2$ where $\alpha$ monotonically decreases
from $\alpha \approx 1$ to $\alpha \approx 0.4$; the value $\alpha \approx 0.4$ for large $\delta \approx 5.0$ lies well in the range
$0.3 < \alpha < 0.5$ and has been obtained for the nearest neighbor FPU model \cite{TwoFifth} in the past.

Actually, such a picture has been proposed for both classical \cite{XYCelia3Regions} and quantum \cite{ThreeRegimesQnt} long-ranged interacting
models (also see \cite{LR_Mukamel}). However, one needs more detailed simulation in each region to ascertain this accurately.

\begin{figure}[htb]
\centerline
{\includegraphics[width=5cm,angle=-90]{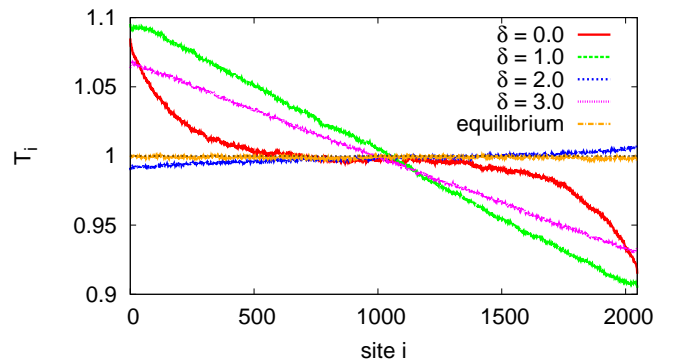}}
\caption{(Color online) Temperature profiles for $N=2048$ and different values of $\delta$; $T_0 = 1.0$ and $\Delta T = 0.2$. For the
curve marked `equilibrium' $\Delta T = 0$ and $\delta = 2.0$.}
\label{fig:TP}
\end{figure}

Next, the local temperature profile $T_i$ is shown for different values of $\delta$ and $N = 2048$ in Fig. \ref{fig:TP}. For $\delta = 0.0$, the
temperature profile is highly nonlinear and becomes less nonlinear as $\delta$ is increased, except for $\delta=2.0$, for which the temperature
profile is nearly flat (in fact, it has a small slope in the $wrong$ sense for reasons unclear to us). The curve marked `equilibrium' in
Fig. \ref{fig:TP} corresponds to $\Delta T = 0$ and $\delta = 2.0$. This is to demonstrate that the long-range interacting system equilibrates
properly ($T_i = T_0 = 1.0$) in absence of external thermal drive at $\delta = 2.0$. 
Evidently, the local slope (say, around $i = N/2$) of the temperature profile also changes non-monotonically as $\delta$ is varied with a nearly flat
profile for $\delta = 2.0$, again similar to the coupled harmonic oscillators.

To summarize the results presented in this section, we find that $\delta = 2.0$ has a very special behavior which leads to a non-monotonic
variation of various quantities such as the conductivity, the slope of the temperature profiles in the bulk of the system, and the divergence
exponent $\alpha$. For $\delta = 2.0$, our long-range interacting nonlinear system exhibits transport properties similar to an integrable
model. In the next section, we provide additional numerical evidence to demonstrate that the results obtained here are physical and not an
artifact of the nonequilibrium setup e.g., the model of heat baths, boundary conditions, interaction potential of the leads, etc.

\section{Results from equilibrium molecular dynamics}
\label{equi}
In order to understand the results obtained in the previous section, we next study the equilibrium (microcanonical) version of the LR-FPU
model using equilibrium molecular dynamics simulation. We remove the baths (along with the leads) and close the 1D chain to obtain a periodic
lattice where each oscillator interacts with all the others according to Eq. (\ref{H}). The total energy of the system is a conserved quantity
now, which is achieved by rescaling the momenta $p_i$ appropriately after they are assigned randomly at $t = 0$. We choose the energy density
$u \equiv \langle \mathcal{H}\rangle/N = 1.0$ which is roughly the same as the average energy of each oscillator at $T_0 = 1.0$ (the average
temperature for which we performed the nonequilibrium simulations).

First, we compute the largest Lyapunov exponent $\lambda_{max}$ as a function of the parameter $\delta$. The numerical computation of $\lambda_{max}$
here has been performed using the algorithm by Benettin {\it et al.} \cite{Benettin}. Note that, the larger the value of $\lambda_{max}$ the greater
is the chaoticity of the model and vice-versa. For integrable systems, such as the harmonic oscillators, $\lambda_{max} = 0$.
\begin{figure}[htb]
{\includegraphics[width=5cm,angle=-90]{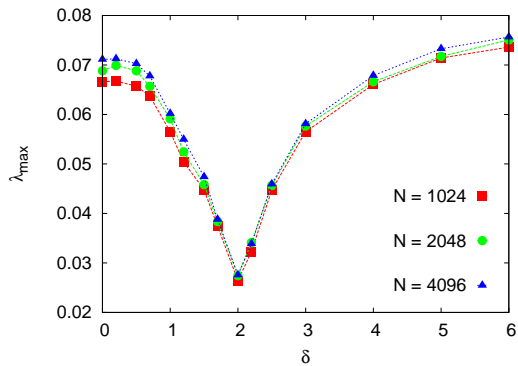}}
\caption{(Color online) Variation of the largest Lyapunov exponent $\lambda_{max}$ with $\delta$ for different system size $N$.
The energy density is set to unity.}
\label{fig:LLExp}
\end{figure}

We show the largest Lyapunov exponent $\lambda_{max}$ as a function of the parameter $\delta$ in Fig. \ref{fig:LLExp}. As can be clearly
seen there is a nonmonotonicity in the behavior of $\lambda_{max}$ exactly at $\delta = 2.0$ for which we have obtained nonmonotonic behavior
for $\kappa$ from nonequilibrium simulations (Fig. \ref{fig:Curr_a}).
For fixed $N$, $\lambda_{max}$ has a minimum at $\delta = 2.0$, implying a suppression of chaos. We speculate that this lack of chaoticity
pushes the nonlinear long-ranged system towards an integrable limit which, in turn, is responsible for the nearly ballistic transport features
($\kappa \sim N$ and negligible temperature gradient) at $\delta = 2.0$ under nonequilibrium conditions. Since in ballistic transport the heat
carriers propagate with minimum scattering (least resistance), the thermal conductivity is a maximum at $\delta = 2.0$, as was obtained in the
previous section.

\begin{figure}[htb]
{\includegraphics[width=5.5cm,angle=-90]{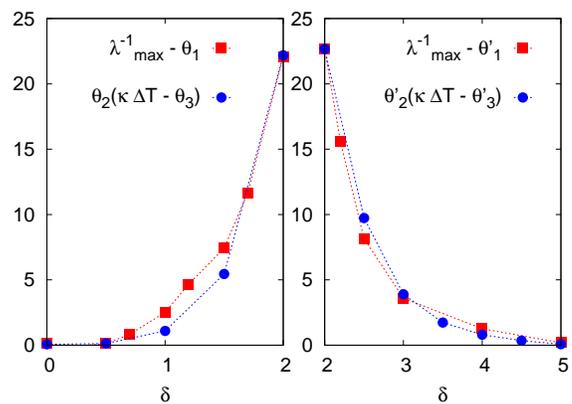}}
\caption{(Color online) Comparison of the variation of the inverse of the largest Lyapunov exponent $\lambda_{max}^{-1}$ and conductivity
$\kappa$ with $\delta$ for $N = 2048$. The curves for $\lambda_{max}^{-1}$ and $\kappa$ have been shifted and rescaled suitably by the real
numbers $(\theta_1,\theta_2,\theta_3)$ for $0 < \delta < 2$ and $(\theta'_1,\theta'_2,\theta'_3)$ for $\delta > 2$ .}
\label{fig:Lyap_kappa}
\end{figure}

Let us extend the connection between $\lambda_{max}$ and $\kappa$ further.
In Fig. \ref{fig:Lyap_kappa}, we have shown $\lambda_{max}^{-1}$ (obtained from Fig. \ref{fig:LLExp}) and $\kappa$ (from Fig. \ref{fig:Curr_a})
by shifting and rescaling these two quantities as $(\lambda^{-1}_{max} - \theta_1)$ and $\theta_2 (\kappa \Delta T - \theta_3)$; $\theta$'s are
real numbers. It can be seen that in both the regions $0 \le \delta < 2$ and $\delta > 2$, the two quantities follow each other approximately.
Thus using the equilibrium results ($\lambda_{max} ~vs.~ \delta$) one can understand the nonequilibrium results ($\kappa ~vs.~ \delta$) qualitatively
and, to an appreciable extent, quantitatively.

Another set of numerical experiments can be performed by exploiting the connection between thermal transport and energy diffusion
since both essentially describe the same microscopic process. It is well established that for normal and ballistic heat transport
the corresponding energy diffusion should also be normal and ballistic respectively. In recent times a lot of effort has been dedicated
to extend such a connection to the regime of anomalous heat transport (see \cite{NJPSpread,Espread}, and references therein). In the
following, we study energy diffusion in our equilibrium model by computing the spatiotemporal excess energy correlation function using
the relation (for a microcanonical system) \cite{FC,Methods}
\begin{equation}
\rho_E(r,t) = \frac{\langle \Delta E_j(t) \Delta E_i(0) \rangle}{\langle \Delta E_i(0) \Delta E_i(0) \rangle} + \frac 1 {N_b - 1},
\label{rho_E}
\end{equation}
where we coarse grain the lattice as $N_b = N/b$, $r = (i-j)b$ and $b$ is the number of sites (oscillators) in each bin and 
$\Delta E_k = E_k - \langle E_k \rangle$ is the excess energy of the $k-th$ bin ($1 \le k \le N_b$) \cite{Methods}.

\begin{figure}[htb]
\centerline
{\includegraphics[width=5.5cm,angle=-90]{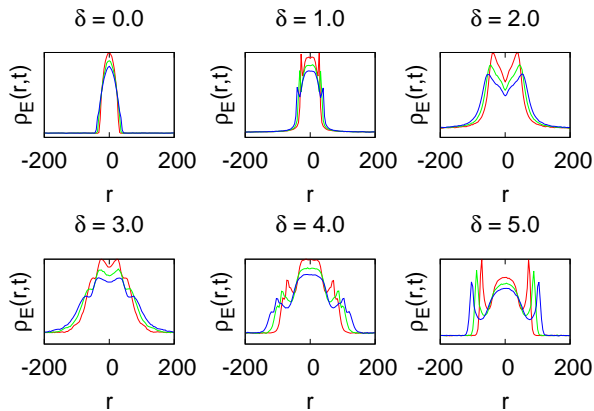}}
\caption{(Color online) Evolution of the excess energy correlation function $\rho_{_E}(r,t)$ for different values of $\delta$.
The three curves (from top to bottom in each box) for each value of $\delta$ correspond to times $t = 1000,1200,1400$. Here $N = 2048$ and energy
density $u = 1.0$.}
\label{fig:ESpread}
\end{figure}

For an equilibrated system, we calculate $\rho_{_E}(r,t)$ as a function of $\delta$ for a fixed $N = 2048$ (we set $b = 4$) for different values
of time after the system has attained equilibrium. The evolution of $\rho_E(r,t)$ at times $t = 1000, 1200,1400$ for different values of $\delta$
is shown in Fig. \ref{fig:ESpread}. Note that, quite surprisingly, the speed of propagation is slower for $\delta = 0.0$ ($\rho_{_E}(r,t)$ is quite
localized around $x = 0$) than for, say, $\delta = 5.0$ (for which the long-range model approaches the nearest neighbor FPU model).
The spatiotemporal distributions $\rho_{_E}(r,t)$ for different times $t$ can be collapsed by rescaling the curves as
$t^{\gamma} \rho_{_E}(r t^{-\gamma}, t)$ where $\gamma < 1/2, \gamma = 1/2, \gamma > 1/2,$ and $\gamma = 1$ imply
sub-diffusion, normal diffusion, super-diffusion and ballistic spreading respectively. 
The exponent $\gamma$ is related to the exponent $\beta$, governing the temporal variation of the mean square deviation (MSD)
$\langle \Delta r^2 (t)\rangle_{_E} \sim t^{\beta}$, as $\beta = 2 \gamma$. Thus when heat transport is ballistic $\beta = 2$.

\begin{figure}[htb]
\centerline
{\includegraphics[width=5.5cm,angle=-90]{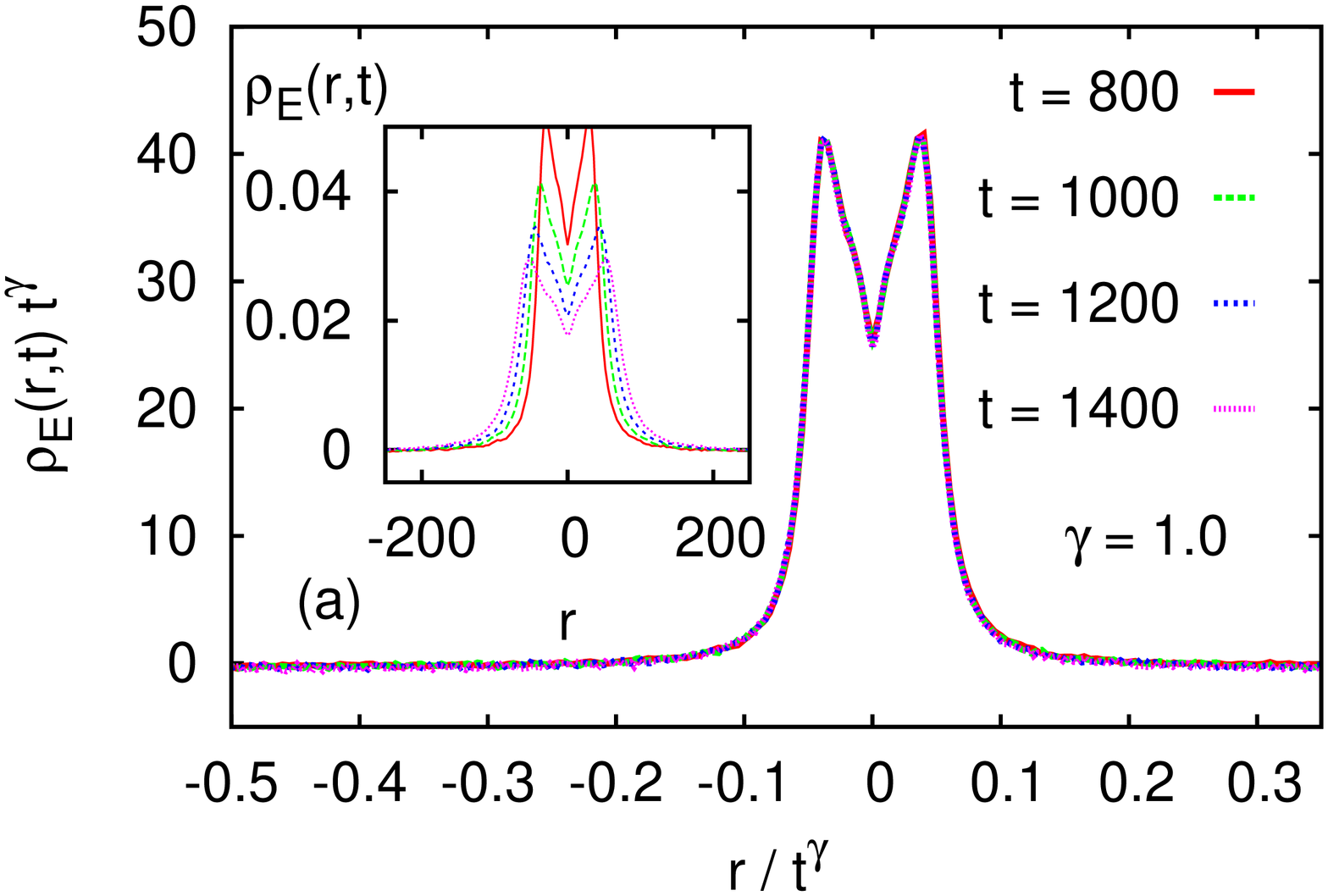}}
{\includegraphics[width=5.5cm,angle=-90]{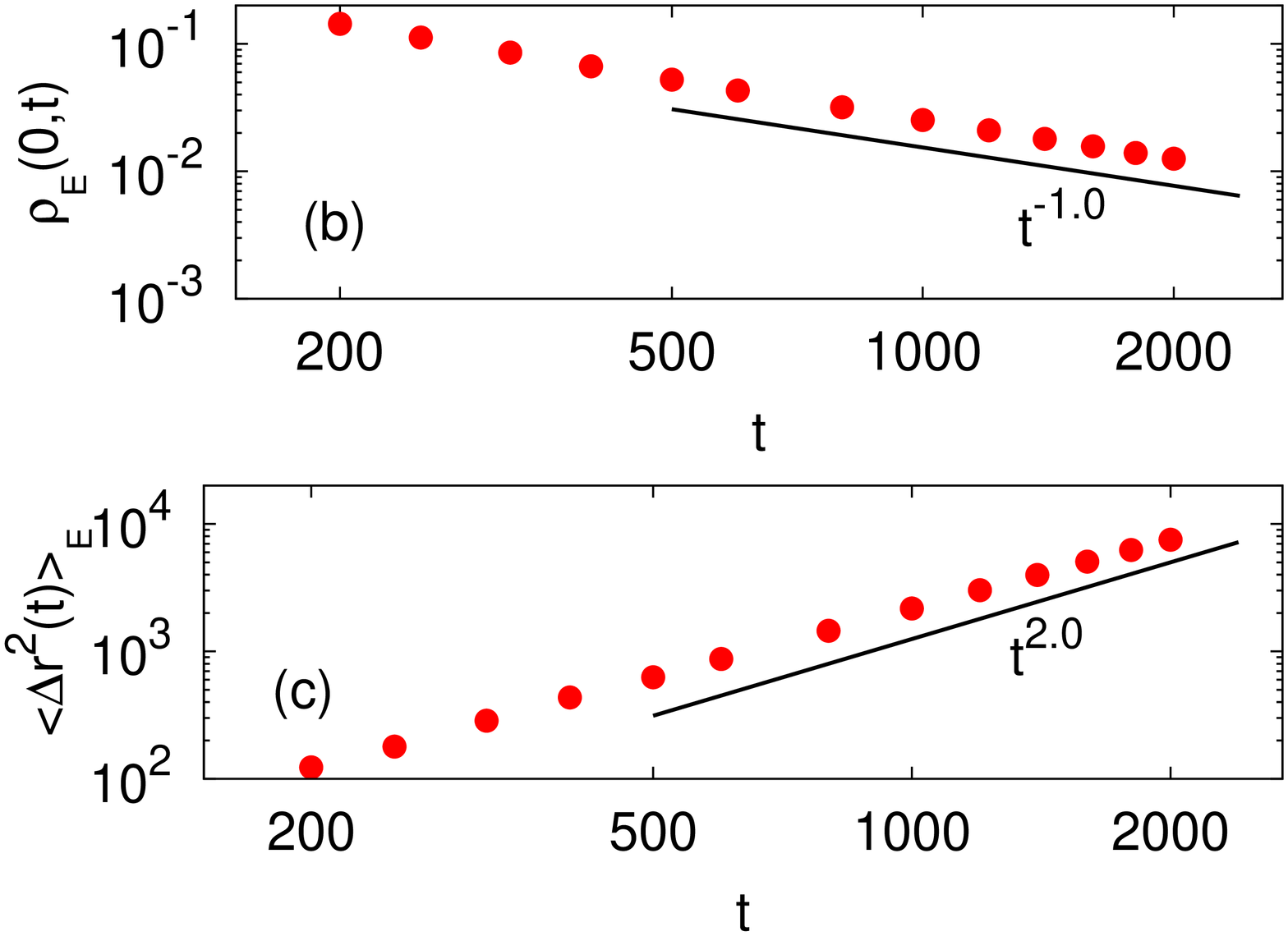}}
\caption{(Color online) (a) Collapse of excess energy correlation functions $\rho_{_E}(r,t)$ for $\delta  = 2.0$ for different
times $t = 800,1000, 1200, 1400$ collapse when rescaled as $t^{\gamma} \rho_{_E}(r t^{-\gamma}, t)$ with $\gamma = 1.0$. The inset
shows the same curves without the rescaling. Variation of (b) $\rho_{_E}(0,t)$ and (c) MSD $\langle \Delta r^2 (t) \rangle_{_E}$
with time $t$. As before, $N = 2048$ and $u = 1.0$.}
\label{fig:A20}
\end{figure}

From Fig. \ref{fig:A20}(a) we find that for $\gamma = 1.0$ one obtains an excellent collapse of the rescaled functions
$\rho_{_E}(r, t)$ shown in the inset for different times $t = 800,1000,1200,1400$. Thus $\beta = 2\gamma = 2.0$ and therefore one
should expect a ballistic spread of energy. In Fig. \ref{fig:A20}(b) we exhibit the function $\rho_{_E}(r = 0, t)$ that demonstrate
that such a scaling (with $\gamma = 1$) persists up to the largest times $t = 2000$ that we have simulated. We also compute the MSD
of the excess energy distribution \cite{NJPSpread}
\begin{equation}
\langle \Delta r^2 (t) \rangle_{_E} = \sum_{r = - N/2}^{N/2} r^2 \rho_{_E}(r, t).
\label{msd}
\end{equation}
The variation of MSD with time $t$ is particularly useful to quantify the speed and identify the nature of energy diffusion. This
is shown in Fig. \ref{fig:A20}(c) for the LR-FPU model at $\delta = 2.0$. As can be seen, we obtain 
$\langle \Delta r^2 (t) \rangle_{_E} \sim t^\beta$ with $\beta = 2.0$ indicating ballistic propagation. All these results strongly
suggest that the ballistic-like heat transport (marked by the nearly linear divergence of the conductivity with system size $N$) as
was seen in Fig. \ref{fig:kN} for $\delta = 2.0$ is a physical property of our model.
For other values of $\delta$ depicted in Fig. \ref{fig:ESpread}, we estimate that $\gamma$ is in the range $3/5 < \gamma < 3/4$, thus
implying super-diffusion and therefore violation of Fourier's law in nonequilibrium, as we expect from Fig. \ref{fig:kN}(a). One can also
compute the exponents $\beta$ and relate it to the $\alpha$ using the so called {\it connection relations} \cite{Rel1, Rel2} to verify
if they hold. A recent study however argued that such a connection between thermal transport and energy diffusion may not exist in
general for anomalous transport \cite{NoConn}.

\section{Conclusion}
\label{conclusion}
In summary, we have studied thermal transport properties of the Fermi-Past-Ulam model in the presence of long-ranged interactions
that decay with (shortest) distance between the lattice sites as a power-law $|i-j|^{-\delta}$. By attaching two Langevin heat
baths at the two ends of the open chain we compute the heat current, conductivity, and the temperature profiles of the long-ranged
FPU system for different values of the exponent $\delta \ge 0$ using large-scale nonequilibrium molecular dynamics simulations. 
We find that $\kappa$ exhibits an interesting non-monotonic $\delta$ dependence with a maximum at $\delta = 2.0$. From the system
size dependence of $\kappa$, we obtain a power-law divergence $\kappa \sim N^{\alpha}$, where $1/3 \lesssim \alpha \lesssim 1$, for
all values of $\delta$ (violation of Fourier's law); for $\delta = 2.0$ we obtain $\alpha \approx 1$ and the temperature profile has 
a negligible (wrong) slope.

In order to explain these results, we next look into the equilibrium dynamics of the long-ranged model (by removing the heat baths).
We compute the largest Lyapunov exponent $\lambda_{max}$ as a function of $\delta$ and obtain a nonmonotonic dependence that enabled
us to understand the $\delta$ variation of the conductivity $\kappa$, both qualitatively and, to some extent, quantitatively.
The near ballistic divergence of $\kappa$ at $\delta = 2.0$ is explained by the suppression of chaos for the same  $\delta$ value in
the equilibrium simulations. Thus, although the range of interaction changes monotonically with the long-range parameter $\delta$, it is
intriguing to see non-monotonicity in all the quantities we have computed namely, $\kappa$, $\alpha$, and $\lambda_{max}$.

From the computation of the spatiotemporal excess energy correlation function, we confirmed the almost linear divergence of
$\kappa$ for $\delta = 2.0$, which corresponds to a ballistic energy diffusion in the equilibrium model.
Thus for $\delta = 2.0$, from both the nonequilibrium and equilibrium simulations, we find that the nonlinear long-ranged interacting FPU 
model behaves similar to a {\it nearly integrable system} \cite{NIM}. This is quite exciting, since integrable nonlinear long-ranged models
are not encountered very frequently (one such example is the many-body Calogero-Moser system \cite{CM} which is nonlinear, long-ranged, and
exactly integrable) and therefore this needs to be explored further. At this point, it is also quite tempting to recall the connection between
the Fermi-Pasta-Ulam model and the nonlinear yet integrable Toda model \cite{Toda3}, both with nearest-neighbor interactions.

Note that we cannot completely rule out finite size effects in the results presented here which therefore may change as $N$ is increased
further (for larger $N > 16384$, performing numerical calculations become computationally impractical). Finite size effects have plagued several
works even for the short-ranged FPU model in the past and in recent times \cite{FiniteSize}. Nevertheless, we believe that the results reported
here, besides their academic interest, will also be important from the experimental perspective where one deals with finite sized systems at
finite temperatures.

There are a number of possible open questions that we believe are important for this model (or similar models) but have not been addressed in
this paper. This includes analyzing the temperature dependences of conductivity, the generality of these results by studying other classes of
interaction potentials, the presence of disorder and on-site potentials, the existence of discrete breathers that often affect heat transport
\cite{DB3}, extension to higher dimensions etc. We plan to address these issues in the future. Of course any analytical result along these
lines will be extremely desirable to obtain. We hope that the results presented here will encourage further investigations of the heat transport
properties of long-range interacting systems in the future.

{\bf Acknowledgments:} The author gratefully acknowledges fruitful discussions with A. Dhar. This work has been funded by the John Templeton
Foundation.


\end{document}